\documentclass[12pt]{article}

\usepackage{amsfonts}

\usepackage{amssymb}

\begin{document}

\begin{titlepage}

\begin{flushright}

CERN-TH/2000-008\\ LAPTH-775/2000 \\ hep-th/0001178 

\end{flushright}

\vspace{.5cm} 

\begin{center}

{\Large\bf Representations of (1,0) and (2,0) superconformal 
algebras in six dimensions: massless and short superfields}\\ 
\vfill {\large  Sergio Ferrara$^\dagger$ and Emery 
Sokatchev$^\ddagger$ }\\ \vfill  \vspace{6pt} $^\dagger$ CERN 
Theoretical Division, CH 1211 Geneva 23, Switzerland 
\\ \vspace{6pt}

$^\ddagger$ Laboratoire d'Annecy-le-Vieux de Physique 
Th\'{e}orique\footnote[1]{UMR 5108 associ{\'e}e {\`a} 
 l'Universit{\'e} de Savoie} LAPTH, Chemin
de Bellevue - BP 110 - F-74941 Annecy-le-Vieux Cedex, France

\end{center}

\vfill

\begin{center}

{\bf Abstract} 

\end{center}
{\small We construct unitary representations  of (1,0) and (2,0) 
superconformal algebras in six dimensions by using superfields 
defined on harmonic superspaces with coset manifolds $U\hskip-2pt 
Sp(2n)/[U(1)]^n$, $n=1,2$.\\ In the spirit of the $AdS_7/CFT_6$ 
correspondence massless conformal fields correspond to 
``supersingletons" in $AdS_7$. By tensoring them we produce all 
short representations corresponding to $1/2$ and $1/4$ BPS anti-de 
Sitter bulk states of which ``massless bulk" representations are 
particular cases.}

\end{titlepage}

\section{Introduction}

Superconformal field theories in space-time dimensions $d\leq 6$ 
have received a lot of attention in recent time because of their 
connection to $(d-1)$-branes and their near-horizon $AdS_{d+1}$ 
geometries \cite{AGMOO}. The most popular examples are IIB string 
theory $D3$-branes and M-theory five- and two-branes related to 
$d=4,6$ and $3$ dimensional superconformal field theories.

A classification of a certain type of UIR's (called highest-weight 
UIR's) of their algebras has been made in the literature in a 
variety of ways, either by using the oscillator method 
\cite{gm}-\cite{GNW} which is directly linked to the $AdS$ 
interpretation or by using superconformal fields defined on 
$\tilde M_d = \partial AdS_{d+1}$, i.e. ``Minkowski space" 
regarded as the boundary at infinity of anti-de Sitter space.

The second approach has recently been employed \cite{AFSZ,FS} for 
all four-dimensional superconformal algebras $SU(2,2/N)$ to 
construct ``massless" and ``short" representations. The latter are 
the generalization of the ``chiral superfields" of $N=1$ 
supersymmetry \cite{fwz}. The generalization of ``chirality" to 
$N>1$ theories, called ``Grassmann analyticity" \cite{GIO} is made 
transparent by using superfields augmented with ``harmonic 
variables" \cite{GIKOS,hh}, i.e. coordinates of coset spaces 
obtained as quotients $G/H$ where $G$ is the R-symmetry group and 
$H$ is a maximal subgroup of $G$ (with rank $G$ = rank $H$). The 
most convenient choice is to take $H$ to be the maximal torus, 
i.e. the group related to the Cartan subalgebra 
$[U(1)]^{\mbox{\scriptsize rank }G}$ of $G$. Such cosets are 
called ``flag manifolds" \cite{Knapp,hh} and an important property 
is that highest-weight UIR's of $G$ defined on such manifolds 
correspond to ``analytic functions" with some degree of 
homogeneity. For the algebras $SU(2,2/N)$ such manifolds are the 
cosets $SU(N)/[U(1)]^{N-1}$. 

The superfield realization of representations of superconformal algebras has
the important property that chiral or Grassmann analytic superfields form a ring
under multiplication. This property allows one, for example, to obtain all short
representations by straightforward multiplication of analytic superfields.

In the present paper we extend the analysis to the $N=(n,0)$ 
($n=1,2$) conformal supersymmetry in six dimensions, i.e. to the  
superalgebras $OSp(8^*/2n)$. In this case the flag manifolds in 
question are $U\hskip-2pt Sp(2n)/[U(1)]^n$. Accordingly, 
superconformal fields will be defined on ``harmonic superspaces" 
with space-time coordinates $x^\mu$ ($\mu=0,1,\ldots,5$), odd 
coordinates $\theta^\alpha_i$ (left-handed spinors of $SU^*(4)\sim 
SO(5,1)$ and spinors of $U\hskip-2pt Sp(2n)$) and coordinates 
$u^I_i$ on the flag manifold. \footnote{A harmonic superspace of 
this type has been used in \cite{Howe} to describe the $N=(2,0)$ 
tensor multiplet. Our formulation of the same multiplet in Section 
4 differs in the choice of the harmonic coset.} A particular 
output of our investigation will be the superfields describing 
massless conformal fields (``supersingletons" in the $AdS$ 
language \cite{ff2}) which satisfy Dirac-type equations 
\begin{equation}\label{0}
  \partial^{\alpha\alpha_1} 
\omega_{(\alpha_1\alpha_2\ldots\alpha_k)} = 0 
\end{equation}
(or $\square \omega=0$ if $k=0$) with $\partial^{\alpha\beta} = 
\partial_\mu (\gamma^\mu)^{\alpha\beta}$ and $\omega$ totally symmetric 
in their spinor indices, i.e. belonging to the $(0,0,k)$ 
representation of $SU^*(4)$.

We will show that supersingleton superfields are of two kinds: (i) 
those whose first component carries any spin label of the above 
type but is an $U\hskip-2pt Sp(2n)$ singlet; (ii) those which have 
an analytic structure in harmonic superspace and are 
``ultrashort"; their first component is a Lorentz scalar but 
carries  $U\hskip-2pt Sp(2n)$ indices. By tensoring the second 
kind of superfields we are able to produce ``short 
representations" which do not depend on one half or on one quarter 
of the odd variables. In the AdS bulk language \cite{FF}, these 
states correspond to $1/2$ or $1/4$ BPS states, respectively. A 
particular example of such states are the so-called ``massless 
bulk states" which correspond to tensoring two massless 
multiplets. Agreement is found for the classification of such 
states as compared to the ``oscillator method" \cite{GT}.

Massive towers corresponding to $1/2$ BPS states are the K-K 
states \cite{GNW} coming from compactification of M theory on  
$AdS_7\times S_4$ \cite{PPN}-\cite{MTN}. The description of such 
K-K states in terms of the $(2,0)$ superconformal field theory was 
considered by several authors \cite{AOY}-\cite{ ABS}. Extension 
of the analysis to $(1,0)$ theories was also investigated 
\cite{FKPZ,GP}.

The paper is organized as follows. In section 2 we list the 
six-dimensional notations and conventions. In section 3 we recall 
how massless conformal supermultiplets (supersingletons) are 
described by constrained superfields in ordinary superspace. In 
section 4 most of these multiplets are reformulated in harmonic 
superspace and it is shown that two of them are ``ultrashort". In 
section 5 the ``short representations" of the $N=(1,0)$ and 
$(2,0)$ superalgebras are constructed by tensoring the basic 
multiplets.

\section{Notations}

We use the six-dimensional notations of Ref. \cite{ckvp} with some 
minor modifications. The six-dimensional superspace has 
coordinates
\begin{equation}\label{1}
  x^{\alpha\beta}=- x^{\beta\alpha} = x^\mu\gamma_\mu^{\alpha\beta} 
\;, \qquad \theta^\alpha_i\;.
\end{equation}
Here $\alpha, \beta$ are right-handed\footnote{Left-handed spinors 
are denoted, e.g., $\psi_\alpha$ which makes transparent the 
meaning of the contraction $\theta^\alpha\psi_\alpha$.} chiral 
spinor indices of $SU^*(4)\sim SO(5,1)$ and $i$ is a spinor index 
of $U\hskip-2pt Sp(2n)\sim SO(2n+1)$  in the case of $N=(n,0)$ 
supersymmetry, $n=1$ or 2. The latter can be raised and lowered 
with the help of the $U\hskip-2pt Sp(2n)$ matrix: 
\begin{equation}\label{2}
  \lambda_i = \lambda^j\Omega_{ji}\;, \qquad \lambda^i = 
\Omega^{ij}\lambda_j\;, \qquad \Omega_{ij}\Omega^{jk} = 
-\delta^k_i\;. 
\end{equation}
We choose $\Omega$ in the following standard form:
\begin{equation}\label{3}
  \left(
  \begin{array}{rr}
    0 & \cal I \\
    -\cal I & 0
  \end{array}
 \right)
  \end{equation}
where $\cal I$ is the $n\times n$ identity matrix. The odd 
coordinates satisfy a Majorana-Weyl pseudoreality condition: 
\begin{equation}\label{4}
  \bar\theta_\alpha^i = \Omega^{ij}\theta_j^\beta c_{\beta\alpha}
\end{equation}
where $c$ is a $4\times 4$ unitary ``charge conjugation" matrix.

The spinor covariant derivatives 
\begin{equation}\label{5}
  D^i_\alpha = {\partial\over \partial \theta^\alpha_i} -{i\over 
2}\theta^{\beta i} \partial_{\beta\alpha} 
\end{equation}
satisfy the supersymmetry algebra
\begin{equation}\label{6}
  \{ D^i_\alpha, D^j_\beta\} = 
i\Omega^{ij}\gamma^\mu_{\alpha\beta}\partial_\mu\;.
\end{equation}

The generators of $U\hskip-2pt Sp(2n)$ $L_{ij}=L_{ji}$ form the 
algebra 
\begin{equation}\label{7}
  [L^{ij},L^{kl}]= \Omega^{i(k}L^{l)j} + \Omega^{j(k}L^{l)i}
\end{equation}
and commute with an $U\hskip-2pt Sp(2n)$ spinor as follows:
\begin{equation}\label{7'}
  [L^{ij},D^k_\alpha]= -\Omega^{k(i}D^{j)}_\alpha
\end{equation}
where the symmetrization has weight 1.

\section{Massless supermultiplets (supersingletons)}

There exist several types of massless multiplets in six dimensions 
corresponding to UIR's of $OSp(8^*/2n)$, $n=1,2$. All of them can 
be described by constrained superfields following closely the 
four-dimensional case \cite{SHST} (see also the analysis of conformal weights for
six-dimensional superfield constraints in Ref. \cite{Park}).

{\sl (i)} The first type only exists in the case $N=(2,0)$ since 
the corresponding superfield $W^{\{ij\}}(x,\theta)$ is 
antisymmetric and {\sl traceless} in the external indices 
({\underline 5 } of $U\hskip-2pt Sp(4)$). It satisfies the 
constraint \cite{HSiT} 
\begin{equation}\label{8}
  D^{(k}_\alpha W^{\{i)j\}}=0\;.
\end{equation}
One can also impose the reality condition
\begin{equation}\label{8'}
  \overline W_{\{ij\}} = \Omega_{ik}\Omega_{jl} W^{\{kl\}}\;.
\end{equation}

Using the spinor derivative algebra (\ref{6}), it is not hard to 
show that this superfield has the following $\theta$ expansion: 
\begin{equation}\label{9}
  W^{\{ij\}} = \phi^{\{ij\}} + \theta^{\alpha\{i}\psi^{j\}}_\alpha 
+ \theta^{\alpha\{i} \theta^{\beta j\}} F_{(\alpha\beta)} + 
\mbox{\small derivative terms }\;. 
\end{equation}
Here one finds 5 scalars $\phi^{\{ij\}}$, 4 left-handed spinors 
$\psi^i_\alpha$ and a  \underline{10} of $SU^*(4)$ 
$F_{(\alpha\beta)} = \gamma^{\mu\nu\lambda}_{(\alpha\beta)} 
F_{\mu\nu\lambda}$ (a self-dual three-form), as well as a few more 
terms containing derivatives of the above fields. These fields 
satisfy massless equations: 
\begin{equation}\label{10}
  \square\phi^{\{ij\}} = 0\;, \quad 
\partial^{\alpha\beta}\psi^i_\beta = 0\;, \quad 
\partial^{\alpha\beta} F_{(\beta\gamma)}=0\;.
\end{equation}
The latter equation implies that the three-form 
$F_{\mu\nu\lambda}$ is the curl of a two-form,
\begin{equation}\label{11}
  F_{\mu\nu\lambda} = \partial_{[\mu}B_{\nu\lambda]}\;.
\end{equation}
One recognizes the content of the on-shell tensor $N=(2,0)$ 
multiplet in six dimensions \cite{HSiT,bsvp}.

It is instructive to give the on-shell counting of degrees of 
freedom in a  six-dimensional massless multiplet. A massless field 
of non-vanishing spin 
$\omega_{(\alpha_1\alpha_2\ldots\alpha_n)}(x)$ describes an irrep 
of $SU^*(4)$ with Dynkin labels $(0,0,n)$. It is subject to the 
Dirac-type field equation 
\begin{equation}\label{msl}
  \partial^{\alpha\alpha_1} 
\omega_{(\alpha_1\alpha_2\ldots\alpha_n)} = 0
\end{equation}
which clearly implies 
$\square\omega_{(\alpha_1\alpha_2\ldots\alpha_n)} = 0$, i.e. the 
field is indeed massless. Thus one can go to the Lorentz frame in 
which the momentum takes the form $p^\mu=(p^0,0,0,0,0,p^0)$. There 
the little group is $SO(4)\sim SU(2)\times SU(2)'$ and an  
$SU^*(4)$ spinor index is decomposed in a pair of $SU(2)$ indices, 
$\alpha = (a,a')$. Then, in the appropriate basis for the gamma 
matrices the operator $p^{\alpha\alpha_1} = 
p^\mu\gamma^{\alpha\alpha_1}_\mu $ in eq. (\ref{msl}) becomes a 
projector onto, e.g., the indices $a'$. This means that the only 
$SU(2)\times SU(2)'$ irreducible part of the multispinor 
$\omega_{(\alpha_1\alpha_2\ldots\alpha_n)}$ surviving in eq. 
(\ref{msl}) is $\omega_{(a_1\ldots a_n)}$, i.e. an $n+1$-plet of 
the first $SU(2)$. The conclusion is that such a field describes 
$n+1$ massless degrees of freedom (in general complex, unless a 
reality condition can be imposed on the field). Applied to eqs. 
(\ref{10}), this counting results in 8 bosons (5 from the scalars 
$\phi^{\{ij\}}$ and 3 from the tensor $F_{(\alpha\beta)}$) and 8 
fermions (from the four spinors $\psi^i_\alpha$). Note that these 
degrees of freedom are real if the reality condition (\ref{8'}) is 
imposed. 

Concluding the discussion of the tensor multiplet we note that the 
$SU^*(4)$ and $U\hskip-2pt Sp(4)$ quantum numbers and the 
dimensions (relative to the first component) of the components 
found in the on-shell superfield expansion (\ref{9}) exactly match 
those of the states of the ``doubleton" supermultiplet listed in 
Table 1 of Ref. \cite{GT} \footnote{Note that, compared to Ref. \cite{GT}, we
use reversed Dynkin labels.}

All other types of massless multiplets exist in both cases 
$N=(n,0)$, $n=1,2$.

{\sl (ii)} The second type is described by a superfield 
$W^i(x,\theta)$ which is in the fundamental UIR of $U\hskip-2pt 
Sp(2n)$. The corresponding constraint is 
\begin{equation}\label{12}
  D^{(k}_\alpha W^{i)}=0\;.
\end{equation}
In the case $N=(1,0)$ the superfield has a very short expansion
\begin{equation}\label{13}
N=(1,0):\qquad   W^i = \phi^i + \theta^{\alpha i}\psi_\alpha + 
\mbox{\small derivative terms }\;. 
\end{equation}
The doublet of scalars $\phi^i$ and the spinor $\psi_\alpha$ 
satisfy the field equations
\begin{equation}\label{14}
  \square\phi^{i} = 0\;, \quad 
\partial^{\alpha\beta}\psi_\beta = 0\;.
\end{equation}
This is the $N=(1,0)$ hypermultiplet \cite{Sohnius} in  six 
dimensions with $2+2$ complex on-shell degrees of freedom (note 
that one cannot impose a reality condition on the superfield 
$W^i$).

In the case $N=(2,0)$ the expansion of $W^i$ becomes
\begin{eqnarray}
  N=(2,0): \qquad W^i &=&   \phi^i + \theta^{\alpha }_j\psi^{[ij]}_\alpha 
 + \theta^\alpha_k \theta^\beta_l \epsilon^{klij}  F_{(\alpha\beta)j} \nonumber\\
  && + \theta^\alpha_j \theta^\beta_k \theta^\gamma_l 
\epsilon^{ijkl}\chi_{(\alpha\beta\gamma)} + \mbox{\small d. t. } 
\label{15} 
\end{eqnarray}
The components are scalars $\phi^i$ ({\underline 4 } of 
$U\hskip-2pt Sp(4)$), spinors $\psi^{[ij]}_\alpha$ ({\underline 
{5} + \underline {1}} of $U\hskip-2pt Sp(4)$), three-forms 
$F^i_{(\alpha\beta)}$ ({\underline 4} of $U\hskip-2pt Sp(4)$) and 
a \underline{20} of $SU^*(4)$ $\chi_{(\alpha\beta\gamma)}$. These 
fields satisfy the massless equations 
\begin{equation}\label{16}
 \square\phi^{i} = 0\;, \quad 
\partial^{\alpha\beta}\psi^{[ij]}_\beta = 
\partial^{\alpha\beta}F^i_{(\beta\gamma)} =
\partial^{\alpha\beta}\chi_{(\beta\gamma\delta)} = 0
\end{equation}
and thus describe 16 bosons (4 from $\phi^i$ and 12 from 
$F^i_{(\alpha\beta)}$) and 16 fermions (12 in $\psi^{[ij]}_\alpha$ 
and 4 in $\chi_{(\alpha\beta\gamma)}$) (all complex). The fields 
in (\ref{15}) match the states of the ``doubleton" supermultiplet 
listed in Table 2 of Ref. \cite{GT}. 
 
{\sl (iii)} The next multiplet stands apart since it is the only 
one described by a superfield without external indices, 
$W(x,\theta)$ (it can be made real, $\overline W = W$). The 
corresponding constraint is second-order in the spinor 
derivatives: 
\begin{equation}\label{17}
 D^{(i}_\alpha D^{j)}_\beta W = 0\;.
\end{equation}

In the case $N=(1,0)$ the superfield expansion is 
\begin{equation}\label{18}
N=(1,0):\qquad    W = \phi + \theta^\alpha_i \psi^i_\alpha + 
\theta^{\alpha i}\theta^\beta_i F_{(\alpha\beta)} + \mbox{\small 
d. t. } 
\end{equation}
where the fields satisfy the massless equations
\begin{equation}\label{20'}
 \square\phi = 0\;, \quad 
\partial^{\alpha\beta}\psi^{i}_\beta = 
\partial^{\alpha\beta}F_{(\beta\gamma)} = 0\;. 
\end{equation}
This is the so-called ``linear multiplet" of Ref. \cite{bsvp} 
describing $4+4$ real (if $W$ is real) degrees of freedom.

In the case $N=(2,0)$ the components of the superfield are
\begin{eqnarray}
  N=(2,0): \qquad W &=&   \phi + \theta^\alpha_i \psi^i_\alpha + \theta^{\alpha 
}_{i}\theta^\beta_j F^{[ij]}_{(\alpha\beta)} + \theta^\alpha_i 
\theta^\beta_i \theta^\gamma_k 
\epsilon^{ijkl}\chi_{(\alpha\beta\gamma)l} \nonumber\\ 
  && + \theta^\alpha_i 
\theta^\beta_j \theta^\gamma_k 
\theta^\delta_l\epsilon^{ijkl}\sigma_{(\alpha\beta\gamma\delta)} + 
\mbox{\small d. t. }  \label{19} 
\end{eqnarray}
and they obey the massless field equations
\begin{equation}\label{20}
 \square\phi = 0\;, \quad 
\partial^{\alpha\beta}\psi^{i}_\beta = 
\partial^{\alpha\beta}F^{[ij]}_{(\beta\gamma)} =
\partial^{\alpha\beta}\chi^i_{(\beta\gamma\delta)} = 
\partial^{\alpha\beta}\sigma_{(\beta\gamma\delta\varepsilon)} =
0\;.
\end{equation}
This amounts to $24+24$ real (if $W$ is real) degrees of freedom. 
The corresponding ``doubleton" supermultiplet of Ref. \cite{GT} is 
listed in Table 3 for $j=1$. 

{\sl (iv)} Finally, there exists a series of multiplets described 
by superfields with external Lorentz indices, 
$W_{(\alpha_1\ldots\alpha_n)}(x,\theta)$ in the  $SU^*(4)$ UIR 
with Dynkin labels $(0,0,n)$. These superfields can be made real 
in the case $n=2k$. Now the constraint takes the form 
\begin{equation}\label{21}
  D^i_{[\beta}W_{(\alpha_1]\ldots\alpha_n)} = 0\;.
\end{equation}
The resulting expansion is 
\begin{equation}\label{22}
N=(1,0):\quad    W_{(\alpha_1\ldots\alpha_n)} = 
\phi_{(\alpha_1\ldots\alpha_n)} + \theta^\beta_i 
\psi^i_{(\beta\alpha_1\ldots\alpha_n)} + \theta^{\beta 
i}\theta^\gamma_i F_{(\beta\gamma\alpha_1\ldots\alpha_n)} + 
\mbox{\small d. t. } 
\end{equation}
describing $(2n+4)+(2n+4)$ massless degrees of freedom or
\begin{eqnarray}
  N=(2,0): &&   W_{(\alpha_1\ldots\alpha_n)} = \phi_{(\alpha_1\ldots\alpha_n)} + 
\theta^\beta_i \psi^i_{(\beta\alpha_1\ldots\alpha_n)} + 
\theta^{\beta}_i\theta^\gamma_j 
F^{[ij]}_{(\beta\gamma\alpha_1\ldots\alpha_n)} \label{23}\\ 
  &&\ \ \ + \theta^\beta_i 
\theta^\gamma_j \theta^\delta_k 
\epsilon^{ijkl}\chi_{l(\beta\gamma\delta\alpha_1\ldots\alpha_n)} + 
\theta^\beta_i \theta^\gamma_j \theta^\delta_k\theta^\varepsilon_l 
\epsilon^{ijkl}\sigma_{(\beta\gamma\delta\varepsilon\alpha_1\ldots\alpha_n)} 
+ \mbox{\small d. t. } \nonumber  
\end{eqnarray}
describing $(8n+24)+(8n+24)$ massless degrees of freedom. The 
corresponding ``doubleton" supermultiplet of Ref. \cite{GT} is 
listed in Table 3 for $j>1$.  

The highest-weight UIR's of the $OSp(8^*/2n)$ algebras will be denoted by 
$$
{\cal D}(\ell,J_1,J_2,J_3;a_1,\ldots, a_n)
$$
where $\ell$ is the conformal dimension, $J_1,J_2,J_3$ are the 
$SU^*(4)$ Dynkin labels and $a_1,\ldots, a_n$ are the $U\hskip-2pt
Sp(2n)$  Dynkin labels of the first component. The analytic 
supersingletons for $n=2$ correspond to ${\cal D}(2,0,0,0;1,0)$ 
and ${\cal D}(2,0,0,0;0,1)$. The other supersingletons correspond 
to ${\cal D}(2+k,0,0,k;0,0)$ for $k=0,1,2,\ldots$. In Section 5.2 
we will show that the short representations corresponding to 
analytic superfields are given by ${\cal D}(2p+4q,0,0,0;p,2q)$ for 
$p,q=0,1,2,\ldots$.

\section{Harmonic superspace formulation}

The massless multiplets {\sl (i-iii)} admit an alternative 
formulation in harmonic superspace \cite{GIKOS,HStT,Howe}. The advantage of 
this formulation is that the constraints (\ref{8}), (\ref{12}) 
become simply conditions for Grassmann analyticity (i.e., 
independence of the superfield of part of the odd coordinates) 
whereas the constraint (\ref{17}) becomes a linearity condition. 
This new simple form of the constraints greatly simplifies the 
tensor multiplication of the multiplets. 

Harmonic superspace is obtained by augmenting ordinary superspace 
$x^\mu,\theta^\alpha_i$ by an internal space (a coset of the 
automorphism group of the supersymmetry algebra). In our case the 
relevant cosets are
\begin{equation}\label{24}
  N=(n,0):\qquad {U\hskip-2pt Sp(2n)\over [U(1)]^n}\;.
\end{equation}
These cosets can be parametrized by harmonic variables forming a 
matrix of $U\hskip-2pt Sp(2n)$: 
\begin{equation}\label{25}
  u\in U\hskip-2pt Sp(2n): \qquad u^I_iu^i_J = \delta^I_J\;, 
\ \ u^I_i \Omega^{ij}u^J_j = \Omega^{IJ}\;, \ \  u^I_i= 
(u^i_I)^*\;. 
\end{equation}
Here the indices $i,j$ belong to the fundamental representation of  
$U\hskip-2pt Sp(2n)$ and $I,J$ are two (four) labels corresponding 
to the $U(1)$ charge(s). The harmonic derivatives (the covariant 
derivatives on the coset (\ref{24})) are the operators 
\begin{equation}\label{26}
  D^{IJ} = \Omega^{K(I}u^{J)}_i{\partial\over\partial u^K_i}\;.
\end{equation}
They are clearly compatible with the defining conditions 
(\ref{25}) and act on the harmonics as follows:
\begin{equation}\label{27}
   D^{IJ}u^K_i = -\Omega^{K(I}u^{J)}_i\;.
\end{equation}
In fact, it is easy to see that these derivatives form the algebra 
of $U\hskip-2pt Sp(2n)$ (see (\ref{7})) realised on the indices 
$I,J$ of the harmonics. In particular, the operators $H=-2D^{12}$ 
in the case $U\hskip-2pt Sp(2)$ and $H'=-2D^{14}$, $H''=-2D^{23}$ 
in the case $U\hskip-2pt Sp(4)$ correspond to the $U(1)$ charges: 
\begin{eqnarray}
 U\hskip-2pt Sp(2): &\quad& Hu^1_i=u^1_i\;, \ Hu^2_i=-u^2_i\;; \label{28'}\\
 U\hskip-2pt Sp(4): &\quad& H'u^1_i=u^1_i\;, \ H'u^2_i=0\;, \ H'u^3_i=0\;, \ H'u^4_i=-u^4_i\;, 
\nonumber \\ 
  &\quad& H''u^1_i=0\;, \ H''u^2_i=u^2_i\;, \ H''u^3_i=-u^3_i\;, \ H''u^4_i=0\;.  
 \label{28} 
\end{eqnarray}
A key ingredient of the harmonic superspace approach of Refs.  
\cite{GIKOS} is the coordinateless realization of cosets like 
(\ref{24}) on harmonic functions homogeneous under the action of 
the charge operators. Such functions are defined by their 
$U\hskip-2pt Sp(2n)$ invariant ``harmonic" expansions. For 
instance, in the case $U\hskip-2pt Sp(2)$ the function $f^1(u)$ 
carries charge $+1$ (like $u^1_i$, see (\ref{28'})) and has the 
expansion 
\begin{equation}\label{29}
 f^1(u)\equiv f^{(+1)}(u) = f^iu^1_i + f^{(ijk)}u^1_iu^1_ju^2_k + 
f^{(ijklm)}u^1_iu^1_ju^1_ku^2_lu^2_m + \ldots\;. 
\end{equation}
In the case $U\hskip-2pt Sp(4)$ the same function $f^1(u)$ carries 
charges $(+1,0)$ (like $u^1_i$, see (\ref{28})) and has the 
expansion    
\begin{equation}\label{30}
  f^1(u)\equiv f^{(+1,0)}(u)  = f^iu^1_i +  g^{(ij)k}u^1_iu^1_ju^4_k + 
h^{ijk}u^1_1u^2_ju^3_k + \ldots\;. 
\end{equation}
In other words, the expansion is formed by the various products of 
harmonics carrying an overall index 1 (i.e., charge $+1$ in the 
case $U\hskip-2pt Sp(2)$ or $(+1,0)$ in the case $U\hskip-2pt 
Sp(4)$). The crucial point about this coordinateless 
parametrization of the coset is that the coefficients in the 
harmonic expansion are manifestly $U\hskip-2pt Sp(2n)$ covariant. 
Another example of a $U\hskip-2pt Sp(4)$ harmonic function is 
\begin{equation}\label{30'}
  f^{12}(u)\equiv f^{(+1,+1)}(u)  = f^{\{ij\}}u^1_iu^2_j +  
g^{(ij)}u^1_iu^2_j +\ldots\;.
\end{equation}
Note the absence of a singlet part (trace) in the coefficient 
$f^{\{ij\}}$, since $\Omega^{ij}u^1_iu^2_j=0$ (see (\ref{25}), 
(\ref{3})). 

As one can see from the above examples, the harmonic functions are 
infinitely reducible under $U\hskip-2pt Sp(2n)$. An important 
point is that the ``step-up" operators (the positive roots) of 
$U\hskip-2pt Sp(2n)$ can be used to impose irreducibility 
conditions on the harmonic functions. In the case $U\hskip-2pt 
Sp(2)$ this is the harmonic derivative $D^{11}$ and in the case 
$U\hskip-2pt Sp(4)$ these are the harmonic derivatives $D^{11}$,  
$D^{12}$,  $D^{13}$,  $D^{22}$. So, for example, 
\begin{eqnarray}
  U\hskip-2pt Sp(2): &\quad& D^{11}f^1(u) = 0  \nonumber\\
     &\quad& \Rightarrow \ f^1(u) = f^iu^1_i \;; \label{31} \\
  U\hskip-2pt Sp(4): &\quad& D^{11}f^1(u) =  D^{12}f^1(u) =  D^{13}f^1(u) =  D^{22}f^1(u) = 0 
\nonumber\\ 
        &\quad& \Rightarrow \ f^1(u) = f^iu^1_i\;; \label{32}\\
&\quad& D^{11}f^{12}(u) =  D^{12}f^{12}(u) =  D^{13}f^{12}(u) =  
D^{22}f^{12}(u) = 0 \nonumber\\ 
        &\quad& \Rightarrow \ f^{12}(u) = f^{\{ij\}}u^1_iu^2_j\;. \label{33}  
\end{eqnarray}
In fact, not all of the conditions (\ref{32}), (\ref{33}) are 
independent, since $D^{11}=2[D^{12},D^{13}]$ and 
$D^{12}=[D^{22},D^{13}]$ (see (\ref{7})). 

Let us now use the $U\hskip-2pt Sp(4)$ harmonics to project the 
defining constraint (\ref{8}) of the $N=(2,0)$ tensor multiplet: 
\begin{equation}\label{34}
   D^{(k}_\alpha W^{\{i)j\}}=0 \ \times 
 \left\{\begin{array}{lll}
    u^1_k u^1_i u^2_j & \Rightarrow & D^1_\alpha W^{12} =0 \\
     u^2_k u^2_i u^1_j & \Rightarrow & D^2_\alpha W^{12} =0
  \end{array}
 \right.\;.
\end{equation}
Here $D^{1,2}_\alpha = D^i_\alpha u^{1,2}_i$ and 
$W^{12}=W^{\{ij\}}u^1_i u^2_j$. In other words, the constraint 
(\ref{8}) now takes the form of a Grassmann analyticity condition:
\begin{equation}\label{35}
D^1_\alpha W^{12} = D^2_\alpha W^{12} =0\;. 
\end{equation}
In addition, the projected superfield $W^{12}$ clearly satisfies 
the $U\hskip-2pt Sp(4)$ irreducibility conditions (\ref{33}). The 
equivalence between the two forms of the constraint follows from 
the obvious properties of the harmonic products $u^1_{[k} u^1_{i]} 
= u^2_{[k} u^2_{i]}=0$ and $\Omega^{ij}u^1_iu^2_j=0$. 

Now it becomes clear that the constraints (\ref{35}) can be solved 
in an appropriate basis in superspace: 
\begin{equation}\label{36}
 D^1_\alpha W^{12} = D^2_\alpha W^{12} =0 \ \Rightarrow \ 
W^{12} = W^{12}(x_A,\theta^1,\theta^2,u)
\end{equation}
where
\begin{equation}\label{37}
  x^{\alpha\beta}_A = x^{\alpha\beta} - 
i\theta^{\alpha(i}\theta^{\beta j)} (u^1_iu^4_j + u^2_iu^3_j)\;, 
\quad \theta^{1\alpha} = \theta^\alpha_4 = \theta^{\alpha}_i 
u^i_4\;, \ \theta^{2\alpha} = \theta^\alpha_3 = \theta^{\alpha}_i 
u^i_3\;. 
\end{equation}
We see that the superfield $W^{12}$ is independent of half of the 
odd coordinates, $\theta^3=-\theta_2$ and $\theta^4=-\theta_1$ 
(hence the name ``Grassmann analytic"). We call such superfields 
``short" (compared to a generic $N=(2,0)$ superfield).

Having solved the constraints (\ref{35}), we should not forget 
that the equivalence with the initial form (\ref{8}) is only 
achieved if the superfield $W^{12}$ satisfies the $U\hskip-2pt 
Sp(4)$ irreducibility conditions (\ref{33}). This is not so 
trivial now, since in the basis (\ref{37}) the harmonic 
derivatives acquire space-time derivative terms: 
\begin{eqnarray}
  D^{11}&=&\partial^{11}+ 
{i\over 4}\theta^{1\alpha}\theta^{1\beta}\partial_{\alpha\beta} \nonumber\\
  D^{12}&=&\partial^{12}+ 
{i\over 4}\theta^{1\alpha}\theta^{2\beta}\partial_{\alpha\beta} 
\label{38}\\ 
  D^{22}&=&\partial^{22}+ {i\over 
4}\theta^{2\alpha}\theta^{2\beta}\partial_{\alpha\beta} 
\nonumber\\
  D^{13}&=&\partial^{13} \nonumber 
\end{eqnarray}
where $\partial^{IJ}$ include the harmonic and $\theta$ partial 
derivatives. Thus, in this basis the $U\hskip-2pt Sp(4)$ 
irreducibility conditions (\ref{33}), 
\begin{equation}\label{33'}
   D^{11}W^{12} = D^{12}W^{12} = D^{13}W^{12} = D^{22}W^{12} =0
\end{equation}
not only eliminate the infinite towers of components in the 
harmonic expansion of $W^{12}$ but also yield the field equations 
on the remaining physical fields. As a result, the analytic 
superfield $W^{12}$ becomes ``ultrashort" (i.e., shorter than a 
generic analytic superfield): 
\begin{equation}\label{39}
  W^{12} = \phi^{12} + {1\over 2}(\theta^{1\alpha}\psi^2_\alpha - 
\theta^{2\alpha}\psi^1_\alpha) + \theta^{1\alpha}\theta^{2\beta} 
F_{(\alpha\beta)} + \mbox{\small d.t.} 
\end{equation}
where $\phi^{12}=\phi^{\{ij\}}(x)u^1_iu^2_j\;$, $\psi^{1,2}_\alpha 
= \psi^i_\alpha(x) u^{1,2}_i$ and 
$F_{(\alpha\beta)}=F_{(\alpha\beta)}(x)$ are the massless fields. 
Thus we recover the component content of eq. (\ref{9}).

Here we should compare our formulation of the $N=(2,0)$ tensor 
multiplet to that of Ref. \cite{Howe}. The harmonic space used 
there is smaller than ours, $U\hskip-2pt Sp(4)/U(1)\times SU(2)$. 
One can easily understand why this is possible \footnote{We thank 
P. Howe for this remark.} by noticing that the superfield $W^{12}$ 
is annihilated not only by all the raising operators (\ref{33'}) 
of $USp(4)$, but also by the lowering operator $D^{24}$. Further, 
the operators $D^{13}$, $D^{24}$ and $D^{14}-D^{23}$ form the 
algebra of $SU(2)\subset U\hskip-2pt Sp(4)$. This effectively 
means replacing one of the factors $U(1)$ in the denominator of 
our harmonic coset $U\hskip-2pt Sp(4)/U(1)\times U(1)$ by $SU(2)$. 
The conclusion is that the supersingleton $W^{12}$ lives on a 
smaller coset. However, in Section 5.2 we are going to tensor 
different realizations of this multiplet, and this is only 
possible if we use the bigger coset $U\hskip-2pt Sp(4)/U(1)\times 
U(1)$.

One can treat the case {\sl (ii)} in the same way. Projecting the 
constraint (\ref{12}) with $u^1_ku^1_i$ we obtain 
\begin{equation}\label{40}
  D^{(k}_\alpha W^{i)}=0 \quad \Leftrightarrow \quad D^1_\alpha 
W^1 = 0\;.
\end{equation}
This constraint of Grassmann analyticity is solved in the 
appropriate basis in $N=(1,0)$ or $N=(2,0)$ superspace and yields
\begin{equation}\label{41}
  D^1_\alpha 
W^1 = 0  \quad \Rightarrow \quad \left\{ 
  \begin{array}{ll}
    W^1 = W^1(\theta^1)\;,  & N=(1,0) \\
     W^1 = W^1(\theta^1,\theta^2,\theta^3)\;,  & N=(2,0)
  \end{array}
 \right. \;.
\end{equation}
In addition, one has to impose the conditions of $U\hskip-2pt 
Sp(2n)$ irreducibility (recall (\ref{31}) and (\ref{32})) 
\begin{eqnarray}
 N=(1,0):  &\ & D^{11}W^1=0\;;\label{42}\\ 
  N=(2,0):  &\ & D^{11}W^1=D^{12}W^1=D^{13}W^1=D^{22}W^1=0\;.\label{43}
\end{eqnarray}
The resulting superfield has the following ``ultrashort" 
expansion: 
\begin{equation}\label{44}
   N=(1,0):  \quad W^1 = \phi^1 + \theta^{1\alpha}\psi_\alpha + \mbox{\small d.t.} 
\end{equation}
with $\phi^1=\phi^i(x)u^1_i$ and $\psi_\alpha=\psi_\alpha(x)$;
\begin{eqnarray}
  N=(2,0): && W^1 = \phi^1 + \theta^{1\alpha}\psi_\alpha -
(\theta^{1\alpha}\psi^{23}_\alpha + \mbox{\small cycle 123}) 
\label{45} \\ 
  && - (\theta^{1\alpha}\theta^{2\beta}F^3_{(\alpha\beta)} + \mbox{\small cycle 123})
+ 6 \theta^{1\alpha}\theta^{2\beta} 
\theta^{3\gamma}\chi_{(\alpha\beta\gamma)}  + \mbox{\small d.t.}  
\nonumber 
\end{eqnarray}
with $\psi^{23}_\alpha = \psi^{\{ij\}}_\alpha(x) u^2_i u^3_j$, 
$F^3_{(\alpha\beta)} = F^i_{(\alpha\beta)}(x)u^3_i$, etc. 
 
Finally, we turn to the case {\sl (iii)} which is different since 
the constraint (\ref{17}) is second-order in the spinor 
derivatives. After projection with $u^I_iu^I_j$ (no summation over 
$I$) it becomes 
\begin{equation}\label{47}
  D^I_\alpha  D^I_\beta  W = 0
\end{equation}
where $I=1,2$ in the case $N=(1,0)$ and $I=1,2,3,4$ in the case 
$N=(2,0)$. This time we do not have Grassmann analyticity but just 
linearity in each of the $\theta^I$. As usual, the superfield $W$ 
also satisfies the $U\hskip-2pt Sp(2n)$ irreducibility conditions. 
Here we only give the expansion of $W$ in the case $N=(1,0)$: 
\begin{equation}\label{47'}
N=(1,0): \qquad  W = \phi  + {1\over 
2}(\theta^{1\alpha}\psi^2_\alpha - \theta^{2\alpha}\psi^1_\alpha) 
+ \theta^{1\alpha}\theta^{2\beta} F_{(\alpha\beta)} + \mbox{\small 
d.t.}  
\end{equation}

\section{Tensoring massless multiplets: products of superfields}

Usually tensoring UIR's is a very non-trivial procedure. The 
problem is to decompose the reducible product into irreps.  The 
interpretation of some of the massless six-dimensional multiplets 
as analytic ($W^{12}$ and $W^1$) or linear ($W$) superfields we 
gave above greatly facilitates this task. We are able to single 
out the principal irreducible part of the various tensor products 
by just imposing our usual harmonic conditions. The tensor product 
of massless fields corresponds to decomposing
in irreducible parts the product of superfields with all of their 
descendants at the same point. Here we restrict ourselves to the rep of lowest
dimension just corresponding to the product of superfields at the same point.

We shall treat the 
cases $N=(1,0)$ and $N=(2,0)$ separately.

\subsection{The case $N=(1,0)$}

As we have shown in eq. (\ref{44}), the superfield $W^1$ is ultra 
short in the sense that its expansion ends at the top ``spin" 
(i.e., $SU^*(4)$ irrep) $(0,0,1)$, as compared to the top spin 
$(0,2,0)$ of a generic $N=(1,0)$ ``long" superfield. Its square 
$(W^1)^2$ still satisfies the same Grassmann and harmonic 
conditions, 
\begin{equation}\label{48}
  D^1_\alpha (W^1)^2 = D^{11} (W^1)^2 =0\;,
\end{equation}
but the content is now different. It is not hard to derive from 
(\ref{48}) that the superfield $(W^1)^2$ has the following 
expansion: 
\begin{equation}\label{49}
  (W^1)^2 = \phi^{11} + \theta^{1\alpha}\psi^1_\alpha + 
\theta^{1\alpha}\theta^{1\beta} A_{[\alpha\beta]} + \mbox{\small 
d.t.} 
\end{equation}
Here we find a triplet of scalars  
$\phi^{11}=\phi^{(ij)}(x)u^1_iu^1_j$, a doublet of spinors 
$\psi^1_\alpha = \psi^i_\alpha(x)u^1_i$ and a vector (i.e., top 
``spin" $(0,1,0)$). All of these fields are off shell and the 
vector is conserved, 
\begin{equation}\label{50}
  \partial^{\alpha\beta} A_{[\alpha\beta]} = 0\;.
\end{equation}
This amounts to $8+8$ {\it off-shell} degrees of freedom. Note 
that unlike $W^1$ itself, the composite superfield 
$W^{11}=(W^1)^2$ can be made real. In the $AdS$ interpretation this is the bulk
multiplet of massless gauge fields.

All higher powers of $W^1$, $(W^1)^p$, $p\geq 3$ are short 
superfields depending on half of the odd variables. Their first 
component is a scalar in the $(p+1)$-plet UIR of $U\hskip-2pt 
Sp(2)$ and the expansion reaches the same top spin $(0,1,0)$. This 
time, however, there are no space-time constraints on the 
components. In the bulk language these states are massive short vector multiplets.

The short superfield $W$ (\ref{47}) is linear in $\theta^{1,2}$, 
therefore its expansion (\ref{47'}) terminates at the top spin 
$(0,0,2)$. The square of $W$, $(W)^2$ satisfies a weaker 
constraint: 
\begin{equation}\label{51}
  (D^I_\alpha)^3(W)^2=0\;, \qquad I=1,2
\end{equation}
which implies that it is bilinear in each $\theta^I$. 
Consequently, the top spin appears in the term 
$\theta^{1\alpha}\theta^{1\beta}\theta^{2\gamma}\theta^{2\delta} 
A_{[\alpha\beta][\gamma\delta]}$, so it is $(0,2,0)$ (and a 
$U\hskip-2pt Sp(2)$ singlet). In fact, this is the maximal spin 
one can have in a generic $N=(1,0)$ 
 ``long" superfield, so in this sense $(W)^2$ is not ``short". Note, 
however, that the top spin in $(W)^2$ is conserved, 
\begin{equation}\label{52}
 \partial^{\alpha\beta}  A_{[\alpha\beta][\gamma\delta]}=0\;,
\end{equation}
whereas this is not the case for any higher power of $W$. The state in (\ref{52})
is a massless bulk graviton while higher powers of $W$ correspond to the massive
graviton recurrences. 

Finally, we have the possibility to tensor $W$ with $W^1$. 
Comparing eqs. (\ref{41}) and (\ref{47}), we see that the product 
$W(W^1)^p$ satisfies the linearity constraint
\begin{equation}\label{53}
  D^1_\alpha D^1_\beta (W(W^1)^p) =0
\end{equation}
as well as the usual harmonic condition
\begin{equation}\label{54}
  D^{11} (W(W^1)^p) =0\;.
\end{equation}
This means that it is linear in $\theta^2$ but the dependence in 
$\theta^1$ is not restricted. Consequently, the top spin appears 
in the term $\theta^{1\alpha}\theta^{1\beta}\theta^{2\gamma} 
\psi^{(p-1)}_{[\alpha\beta]\gamma}$, so it is $(0,1,1)$ (and it 
also is a $p$-plet of $U\hskip-2pt Sp(2)$). The case $p=1$ (i.e., 
the bilinear product $WW^1$) is again special, since the condition 
(\ref{54}) implies that the top spin is conserved, 
\begin{equation}\label{55}
  \partial^{\alpha\beta}\psi_{[\alpha\beta]\gamma} = 0\;.
\end{equation}
In the bulk language the above state corresponds to a massless ``gravitino".

\subsection{The case $N=(2,0)$}

In this case we shall restrict ourselves to the products of 
analytic superfields of the type $W^{12}$ and $W^1$. The 
superfield $W^{12}$ is ultrashort (recall (\ref{39})), its top 
spin being $(0,0,2)$. The square $(W^{12})^2$ still satisfies the 
analyticity constraints (\ref{36}) (as well as the harmonic 
conditions (\ref{33'})), so it only depends on 
$\theta^{1,2}_\alpha$ and its expansion goes up to the top spin 
$(0,2,0)$ found in the term 
\begin{equation}\label{56}
 (W^{12})^2 = \phi^{1122} + \ldots 
+ \theta^{1\alpha}\theta^{1\beta}\theta^{2\gamma}\theta^{2\delta} 
A_{[\alpha\beta][\gamma\delta]} + \mbox{\small d.t.} 
\end{equation}
This is the maximal spin in an analytic $N=(2,0)$ superfield 
depending on two $\theta$'s only. However, this top spin satisfies 
a conservation condition, as is always the case with bilinear 
(current-like) products. As to the higher powers $(W^{12})^p$, 
$p\geq 3$, the top spin there still is $(0,2,0)$ but it is 
unconstrained. 

To summarize, tensoring $W^{12}$'s we obtain the following series 
of UIR's of $OSp(8^*/4)$:
\begin{equation}\label{series1}
  (W^{12})^p = \phi^{\stackrel{\underbrace{\mbox{\scriptsize 1\ldots 1}}}{p}
\stackrel{\underbrace{\mbox{\scriptsize 2\ldots 2}}}{p}} + \ldots 
+  \theta^{1\alpha}\theta^{1\beta}\theta^{2\gamma}\theta^{2\delta} 
A^{\stackrel{\underbrace{\mbox{\scriptsize 1\ldots 1}}}{p-2} 
\stackrel{\underbrace{\mbox{\scriptsize 2\ldots 
2}}}{p-2}}_{[\alpha\beta][\gamma\delta]} + \mbox{\small d.t.} 
\end{equation}
having the first component in the $(p,0)$ and the top spin 
$(0,2,0)$ in the $(p-2,0)$ UIR's of $U\hskip-2pt Sp(4)$.  
\footnote{In the harmonic formalism the representation of 
$U\hskip-2pt Sp(4)$ to which belongs each component is identified 
by just counting the 1's (2's) which gives the number of cells in 
the top (bottom) row of the corresponding Young tableau.} These 
superfields do not depend on one half of the odd variables and 
thus correspond to $1/2$ BPS states in the $AdS$ language. 

Besides $W^{12}$, we have another analytic superfield $W^1$ which 
is ``intermediate short" since it depends on 
$\theta^{1,2,3}_\alpha$ (recall (\ref{41})) (or, to put it 
differently, it does not depend on $1/4$ of the odd variables). It 
is clear that by multiplying the ``short" $W^{12}$'s by 
``intermediate short" $W^1$'s we obtain ``intermediate short" 
composite objects. However, there exists an alternative way of 
constructing such objects \cite{AFSZ,FS}. The choice of harmonic 
projections in eq. (\ref{34}) is not unique. Exchanging, e.g., 2 
with 3 we could obtain another analytic superfield 
$W^{13}(\theta^1,\theta^3)$ which provides an equivalent 
description of the on-shell tensor multiplet. Now, consider the 
product $W^{12}(\theta^1,\theta^2)W^{13}(\theta^1,\theta^3)$. It 
depends on $\theta^{1,2,3}_\alpha$, just like $W^1$. In addition, 
we can impose on it the same harmonic conditions (\ref{33'}) as on 
$W^{12}$ alone. The result is an ``intermediate short" superfield 
with top spin $(0,3,0)$: 
\begin{equation}\label{new}
  W^{12}W^{13} = \phi^{11} + \ldots + \theta^{1\alpha}\theta^{1\beta}
\theta^{2\gamma}\theta^{2\delta} 
\theta^{3\kappa}\theta^{3\sigma}  
A_{[\alpha\beta][\gamma\delta][\kappa\sigma]} + \mbox{\small d.t.} 
\end{equation}
It should be noted that the composite object $(W^1)^2$ has exactly 
the same content. Indeed, it depends on the same $\theta$'s and 
has the same first component (a scalar in the $(0,2)$ of 
$U\hskip-2pt Sp(4)$), so
\begin{equation}\label{equi}
  W^{12}W^{13} \simeq (W^1)^2\; .
\end{equation}

Generalizing the above tensor product we can construct a 
two-parameter series: 
\begin{equation}\label{60}
  (W^{12})^{p+q}(W^{13})^{q}  = \phi^{\stackrel{\underbrace{\mbox{\scriptsize 
1\ldots 1}}}{p+2q} \stackrel{\underbrace{\mbox{\scriptsize 2\ldots 
2}}}{p}} + \ldots +  
\theta^{1\alpha}\theta^{1\beta}\theta^{2\gamma}\theta^{2\delta} 
\theta^{3\kappa}\theta^{3\sigma}  
A^{\stackrel{\underbrace{\mbox{\scriptsize 1\ldots 1}}}{p+2q-2} 
\stackrel{\underbrace{\mbox{\scriptsize 2\ldots 
2}}}{p}}_{[\alpha\beta][\gamma\delta][\kappa\sigma]} + 
\mbox{\small d.t.} 
\end{equation}
having the first component in the $(p,2q)$ and the top spin 
$(0,3,0)$ in the $(p,2q-2)$ UIR's of $U\hskip-2pt Sp(4)$. Note 
that, as usual, the top spin in the bilinear combinations 
$W^{12}W^{13}$ is conserved. These superfields do not depend on 
one quarter of the odd variables and thus correspond to $1/4$ BPS 
states in the $AdS$ language.

We remark that the three bilinear cases $(W^{12})^2$, $W^{12}W^1$ 
and $ W^{12}W^{13} \simeq (W^1)^2$ can be identified with the 
massless (in the $AdS_7$ sense) supermultiplets in Tables 4, 5 and 
6, respectively from Ref. \cite{GT}.

In conclusion we can say that the analytic series $(W^{12})^p$ 
correspond to $1/2$ BPS states in the sense that they preserve 
$1/2$ of the original supersymmetry.  These are the operators 
which classify the K-K states of M-theory on $AdS_7\times S^4$ 
\cite{GNW}. In superfield language  these are the analytic 
superfields in harmonic superspace which do not depend on two of 
the four odd variables in $N=(2,0)$ superspace. The other short 
representations  $(W^{12})^{p+q}(W^{13})^{q}$ with $q>0$ 
correspond to  $1/4$ BPS states. The one with $p=0$, $q=2$ is 
contained in the two-graviton state. 

The above results should be relevant for the analysis of $n$-point 
functions in $D=6$ and the correspondence with $n$-graviton 
amplitudes in M-theory on $AdS_7\times S^4$ \cite{C,B}.

 \section*{Acknowledgements}

We would like to thank M. G\"unaydin and R. Stora for enlightening 
discussions and P. Howe for turning our attention to Ref. 
\cite{Howe}. E.S. is grateful to the TH Divison of CERN for its 
kind hospitality. The work of S.F. has been supported in part by 
the European Commission TMR programme ERBFMRX-CT96-0045 
(Laboratori Nazionali di Frascati, INFN) and by DOE grant 
DE-FG03-91ER40662, Task C.

\vfill\eject

\end{document}